\documentclass{article}
\usepackage[utf8]{inputenc}
\usepackage{graphicx}
\usepackage[T1]{fontenc}
\usepackage[backend=bibtex,style=numeric]{biblatex}
\usepackage[english]{babel}
\usepackage{float}
\usepackage{amsfonts}
\usepackage[utf8]{inputenc}
\usepackage{epsfig}
\usepackage{graphicx}
\usepackage[export]{adjustbox}
\usepackage[linesnumbered,ruled,vlined]{algorithm2e}

\usepackage[fleqn]{amsmath}
\usepackage{subfig}
\usepackage{placeins}

\title{\LARGE \bf
Big Data application in congestion detection and classification using Apache spark 
}

\author{Atousa Zarindast$^{1}$ and Anuj Sharma$^{2}$ %

\thanks{$^{1}$A. Zarindast is PhD student in civil engineering Engineering, Iowa state university, USA}%
\thanks{$^{2}$A. Sharma is professor in civil engineering, Iowa state university, USA}%
}

\begin{document}

\maketitle
\thispagestyle{empty}
\pagestyle{plain}

\begin{abstract}
With the era of big data, an explosive amount of information is now available. This enormous increase of Big Data in both academia and industry requires large-scale data processing systems. A large body of research is behind optimizing Spark's performance to make it state of the art, a fast and general data processing system. Many science and engineering fields have advanced with Big Data analytics, such as Biology, finance, and transportation. Intelligent transportation systems (ITS) gain popularity and direct benefit from the richness of information. The objective is to improve the safety and management of transportation networks by reducing congestion and incidents. The first step toward the goal is better understanding, modeling, and detecting congestion across a network efficiently and effectively. In this study, we introduce an efficient congestion detection model. The underlying network consists of 3017 segments in I-35, I-80, I-29, and I-380 freeways with an overall length of 1570 miles and averaged (0.4-0.6) miles per segment. The result of congestion detection shows the proposed method is 90 \% accurate while has reduced computation time by 99.88 \%.

\end{abstract}

\section{INTRODUCTION}
Given data production speed and different data structures, efficient large scale data processing to analyze and mine patterns is a must. Thereby a number of large-scale data processing frameworks have been developed, such as MapReduce \cite{dean2008mapreduce}, Storm \cite{storm}, TensorFlow \cite{abadi2016tensorflow}. However, while Spark \cite{zaharia2010spark} is a fast and general computing system, all frameworks mentioned above lack generality, as each of them is suitable for certain data computation \cite{tang2020survey}. 
Spark benefits from in-memory data processing. Moreover, as a general system, it supports batch, interactive, and streaming computations. Despite its popularity, Spark has limitations; for instance, it requires a considerable amount of learning and effort to program under the Resilient Distributed Datasets (RRD) programming model.    
RDD abstraction model is a collection of data records partitioned among several computers \cite{zaharia2012resilient}. As compared to MapReduce, Spark is faster, more generic, and flexible. It also supports (standalone or on cluster) computations. However, Spark suffers from poor security, and it has a learning curve. Due to the complexity of the programming model, users must learn and be familiar with APIs before programming their Spark application. Big Data is generally saved and managed in distributed file systems or databases. Hadoop Distributed File System (HDFS) is a highly scalable system that enables running jobs on a cluster. It contains a master node called NameNode and several slaves called DataNodes for storing data. Spark supports HDFS to read/write data from/to HDFS directly.
Spark support R and Python as high-level languages with SparkR and PySpark \cite{pyspark}.  PySpark is a python API for Spark and supports python's lambda function passed to PySpark.
As an efficient data processing system, Spark is used in many application domains such as Genomics, Medicine, Finance \cite{seif2018stock} and Transportation. For instance, Spark-DNAlighning \cite{aljame2019dna} exploits Spark's performance optimization such as broadcast variable, join, partitioning, and in-memory computations for DNA short reads alignment problem. Among other fields, Transportation also gets a direct benefit from an efficient and scalable data processing system. 

Rich in information and enormous in size transportation data could enhance understanding of the transportation network. Moreover, it is an enabler for proactive traffic management systems. 
Traffic network efficiency and safety are among priorities in traffic management and system performance measurement. As efficiency is evaluated in traffic congestion, extensive work is conducted to identify contributing factors in traffic congestion. 
This study utilized the Spark ecosystem to develop a scalable algorithm to detect Spatio-temporal congestion and further identify Recurrent congestion. The study location is Iowa's interstates consisting of I-35, I-80, I-380, and I-29 total of 3017 segments between (0.4-0.6) miles length with a total of 1570 miles length. This sizeable geographical location gives a full view of the network performance, serving as Big Data.

\section{Background}
Intelligent transportation systems (ITS) produce a large amount of data. This enormous input will have a profound impact on transportation networks and ITS. Big Data brings more profit, safety, and efficiency to ITS by designing, planning, and controlling transportation networks.
The purpose of ITS is to provide better service to drivers in transportation systems. ITS incorporates advanced technologies including electronic sensors, data transmissions technologies into transportation network \cite{an2011survey}. 
GPS, sensors, cameras, social media are diverse sources of Terabytes of data in ITS. 
Given the size, complexity, and production speed of data, traditional data processing systems and models are inefficient. Big Data analytic provides a new technical method benefiting a broad range of disciplines \cite{rajabalizadeh2020exploratory}. It can directly benefit from Big Data analytics by resolving data storage, data analysis, and data management problems. Moreover, Big Data analytics can improve real-time traffic operation decisions such as traffic public transportation routing. Traffic engineering covers a broad range of planning decisions, from strategic decisions such as transportation network design \cite{zarindast2018determination,gunay2020multi} to operational such as traffic congestion monitoring \cite{zarindast2019data}.
Big Data can improve safety by analyzing the sensors, predicting incidents, and planning rescue scenarios. Big Data analytics can also improve traffic network efficiency by monitoring real-time traffic congestion and introducing controlling policies. Traffic congestion occurs when the demand for a traffic network exceeds its design capacity. The transportation sector is among the largest carbon-dioxide (CO) emission contributors. Traffic congestion is one of the critical issues that concern the sustainability of traffic networks \cite{rehborn2011empirical}. It contributes to long commuting time, decreases the quality of life, and increase energy consumption. It produces emission and pollution, which is a significant threat to public health \cite{levy2010evaluation}. 
The Spatio-temporal pattern of congestion represents the first step toward developing effective policies for dealing with this issue. Although, sampling error might result in unreliable modeling when scaled \cite{shahhosseini2020improved}. Research in congestion detection literature is limited either spatially for specific zones \cite{a2} or temporally for a short period \cite{zhao2019geographical}. Thereby, it lacks the Spatio-temporal comprehensiveness.

Through careful application of Big Data analytics, we try to reveal a wealth of hidden traffic data insights. We uncover the Spatio-temporal pattern of congestion for the extensive network under study using a distributed programming model. We introduce a data-driven Spatio-temporal congestion detection framework based on (RRD) programming model. Moreover, 
travel delay and cost are considered to be a measure of congestion. We report delay and cost associated with detected congestion patterns.

\section{Data and methodology}
\subsection{Study location}
\begin{figure}[H]
    \includegraphics[scale=0.6]{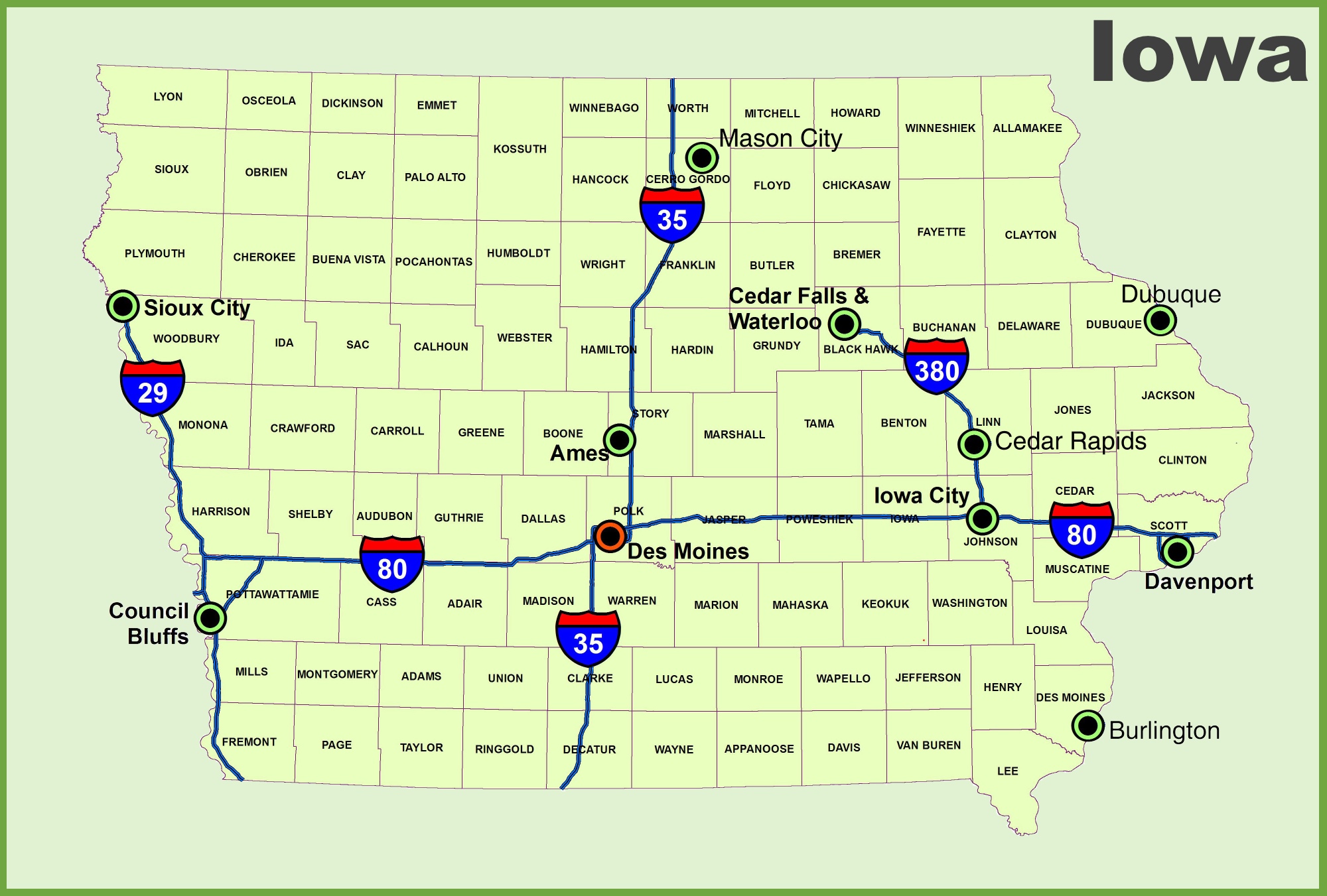}
    \caption{Study region in Iowa, USA}%
    \label{fig:area.}%
\end{figure}

Probe vehicle speed data provided by INRIX \cite{INRIX} with a minute-wise reporting frequency for the entire 2018 calendar year were used in this study. INRIX reports average speed data for each segment in one-minute intervals using probe vehicles' on-board GPS devices. The network of transportation in this study is Iowa's interstates freeways. As shown in Figure \ref{fig:area.} Iowa's freeways includes (I-35, I-80, I-29, and I-380). This network consists of 3017 segments that partition the network almost equal length (from 0.4 to 0.6 miles).

\subsection{Method}
\label{data}
The mathematical abstraction of our data corpus is defined here. Consider the weighted graph of the traffic network denoted by $G=(S, E, W)$, where $S$ defines all nodes of the graph, $E$ denotes the edges, and $W$ denotes the weights of the edges. This study node graphs correspond to consecutive road segments that partition the freeway understudy, and average vehicle speeds $(x)$ in one-minute intervals are reported for each segment. The nodes corresponding to consecutive segments connect weighted edges along the freeway. Since in our study, freeway segments are approximately equal in length (from 0.4 to 0.6 miles), the weight of the nodes in this study was all counted as equal to 1.
The order of road segments describes the connectivity of the graph and the speed values are a (noisy) time-series with the length $N$, for each node $S$ for a given date $v$.
\begin{equation}
\label{x}
    X^v_s=\{x^{t_{1},v},x^{t_{2},v}, \ldots, x^{t_{N},v}\} ,  s\in \{S\} ,  v\in \{D\}
\end{equation}
Where $t_i$ denotes the $ith$ time instant $T=\{\,t_i \mid i \in N\,\}$ and $v$ denotes the date. Overall, we have a third-order tensor $x \in  R_+^ n\times N\times D$\ where $D$ defines the total number of days, $n$ defines the total number of nodes (segments), and $N$ defines the length of the time-series for each date. For example, in this study, since average speed values are being reported in one-minute intervals for each segment the length of the time series $N$ is equal to $24\times 60=1440$. Given our tensor, the major challenge that we faced was the scale of the traffic data. After all, there are 54,000 segments in the entire road network of Iowa that produce 4 gigabytes of daily traffic data.  \\

Zarindast et al. (2020) \cite{a2} proposed a data-driven definition and machine learning algorithm for congestion and congestion patterns. Their proposed framework was tested in Des Moines, Iowa; thereby, the proposed regional parameters perfectly match our network. As a result in this work, inspired by \cite{atousa:online} and based on \cite{a2} important parameters for our algorithm were extracted. These data-driven definitions are the duration of congestion, congestion threshold, and recurrent congestion frequency in the data. We calculated congestion index \ref{eq:CI} for congested regime under data-driven method \cite{a2}. Figure \ref{eq:CI} shows histogram of congestion index for congestion regime with data-driven method \cite{a2}. Congestion index (CI) starts from 0.15; in our framework, we consider the congestion threshold equal to 0.15.
\begin{equation}
CI=\dfrac{RS-S}{RS+S}
\label{eq:CI}%
\end{equation}
\begin{figure}[H]
    \includegraphics[scale=0.9]{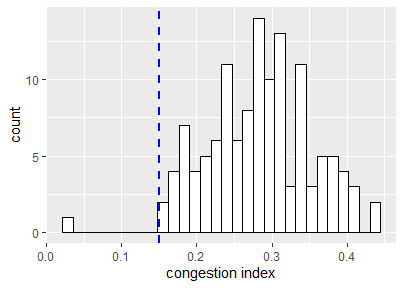}
    \caption{Histogram of congestion index.}%
    \label{fig:CI}%
\end{figure}
   \begin{equation}
       Congestion_i= 
        \begin{cases}
            1 & \text{if $ CI > 0.15 $ for  $\{i \mid$ $i \in (1:1440)  \& $  $[i:i-15] \}$   } \\
            0 & \text{otherwise}
        \end{cases}
    \end{equation}

\begin{equation}
Indicator_i=
\begin{cases}
           1 & \text{if   $\mathlarger{\mathlarger{\sum_{D}}}\dfrac{\sum_{D}{Congestion_i}}{n} > 3$ for  $\{i, D \mid$ $i \in (1:1440) , D \in (1:365)\}$ }\\
        0 & \text{otherwise}
        \end{cases}
\end{equation}
    
\section{Computing machine specifications}
Big Data is generally saved and managed in distributed file systems or databases. Hadoop Distributed File System (HDFS) is a highly scalable system that enables running jobs on a cluster. Spark support read/write from/on HDFS; hence our data storage structure is HDFS. We used PySpark (spark API) to conduct our experiment with the cluster under (RRD) programming model.  The experiments were hosted on-prem Cloudera based high-performance cluster. The cluster was composed of 33 high power machines, including two master nodes and 31 worker nodes.
\\ 
\textbf{Cluster specification:} 
\begin{itemize}
    \item Operating system: Redhat 7.6
    
\item Total RAM:  816 GB
\item Total HDFS Storage: 222 TB
\item Total number of cores: 156
\item Spark Version: 2.3
\item Hadoop Version: 2.6.0
\end{itemize}
The utilized computing power and time for the experiment is described in Table \ref{table:u}.

\begin{table}[H]
    \caption{Utilized processing resources in the algorithm.}
    \label{table:u}

\begin{tabular}{llllll}
\cline{1-4}
Data size (GB) & CPU nodes & Memory (GB)& Computation time (Min) &  &  \\ \cline{1-4}
76.4 GB         & 64                             & 224 GB                           & 10 min                &  &  \\ \cline{1-4}
                &                                &                                  &                       &  &  \\
                &                                &                                  &                       &  & 
            
\end{tabular}
\end{table}
\section{Result}
\begin{figure}[H]
\centering
    \includegraphics[scale=0.7]{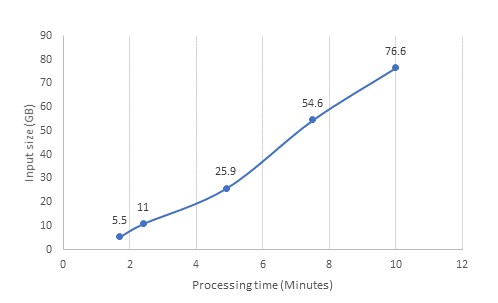}
    \caption{Comparing processing time based on input size. }%
    \label{fig:Processingtime}%
\end{figure}

Figure \ref{fig:Processingtime} compares the resource utilization by increasing the input size. This table can give us an example of how input size if effecting computation times.

\begin{figure}[H]
\centering
    \includegraphics[scale=0.5]{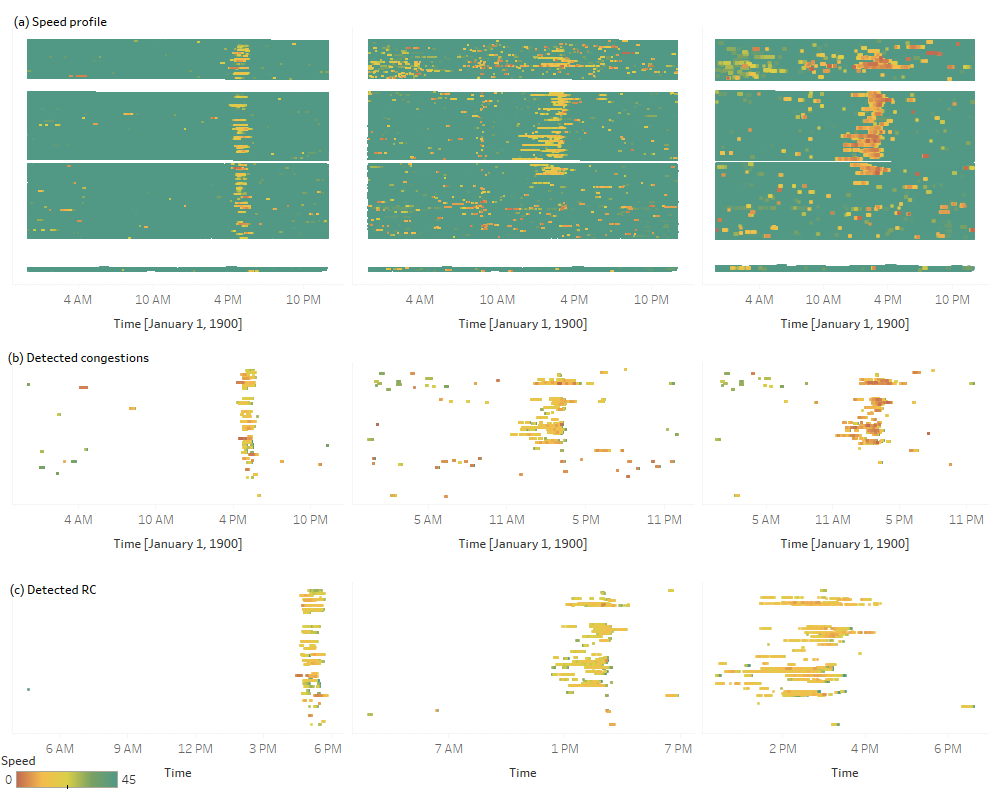}
    \caption{Comparison between (a) actual speed profile, (b) detected congestion, (c) detected RCs for three example segments }%
    \label{fig:RC.}%
\end{figure}

Consider matrix \ref{eq:symmetrical} , in our third-order tensor $x \in  R_+^  n\times N\times D$ (Vector \ref{x}), $S$ is constant for each example segment.It should be noted that speed values are presented in Figure \ref{fig:RC.} only for date and times detected by our algorithm as experiencing an instance of congetion in (b) and RC in (c) for each particular segment.

\begin{equation}
\label{x}
    X^v_s=\{x^{t_{1},v},x^{t_{2},v}, \ldots, x^{t_{N},v}\} ,  s\in \{S\} ,  v\in \{D\}
\end{equation}

\begin{equation}
\label{eq:symmetrical}
   \textrm{Date $v\in \{D\}$}
  \stackrel{\mbox{$N$ Time of day}}{%
    \begin{bmatrix}
    x_{11} &  \cdots & x_{1N} \\
    x_{21}  & \cdots & x_{2N} \\
    \vdots & \ddots & \vdots \\
    x_{3651}  & \cdots & x_{365N}
    \end{bmatrix}}\\
\end{equation}

As shown in Figure \ref{fig:RC.}, our proposed framework could detect Spatio-temporal RCs. These congestion hour periods refer to the morning or evening peak traffic hours.

\begin{figure}[H]
\centering
    \includegraphics[scale=0.7]{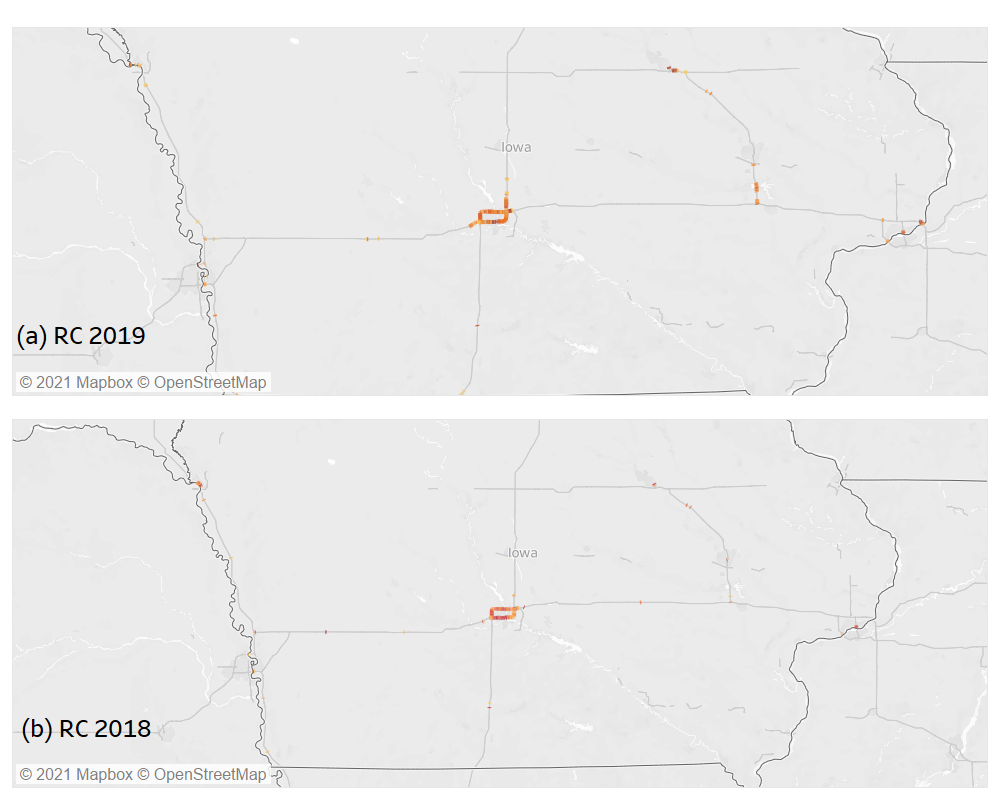}
    \caption{Average RC hours by segment for year (2019)-(a) and year (2018)-(b). }%
    \label{fig:RChours}%
\end{figure}

Figure \ref{fig:RChours} shows average congestion hours by segments. As expected des moine loop experiences RC.

\begin{figure}[H]
\centering
    \includegraphics[scale=0.5]{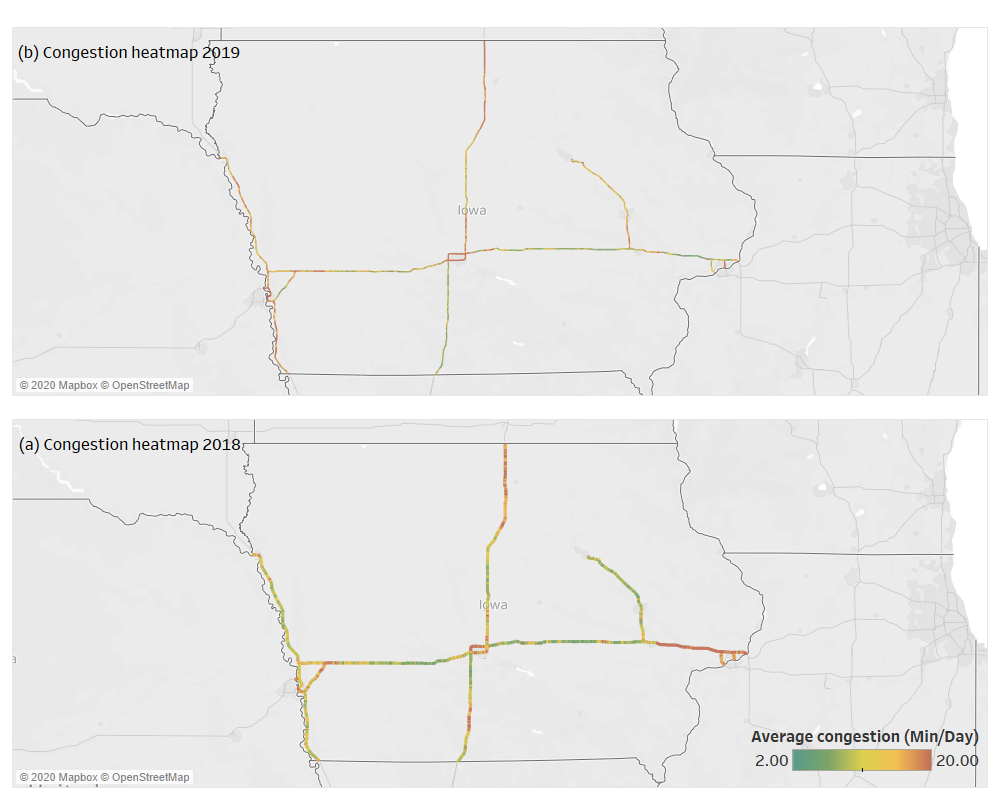}
    \caption{Average congestion hours by segment for year (2019)-(b) and year (2018)-(a). }%
    \label{fig:conhours}%
\end{figure}

Figure \ref{fig:conhours} shows average congestion hours by segments. As expected des moine loop experiences RC.

We investigated the performance of our proposed method with recently developed data driven method \cite{a2} in terms of computation time, computation cost and accuracy. The accuracy of proposed framework has been compared with recently developed data driven method by \cite{a2} using INRIX data for calender year 2016, where study region were Des moine, Iowa, USA. The accuracy has been calculated by comparing the identified congested hours between data driven method proposed by \cite{a2} and developed framework in this study. Considering segment $s$ and date $d$ such as $s\in \{S\} ,  v\in \{D\}$ we have following definitions:
\begin{itemize}
\item True Positive: Congestion has been correctly identified by proposed method $\forall$ $h$, $s$, $d$ (TP)
\item True Negative: Freeflow has been correctly identified by proposed method $\forall$ $h$, $s$, $d$ (TN)
\item False Positive: Congestion has been identified by proposed method incorrectly $\forall$ $h$, $s$, $d$ (FP)
\item False Negative: Freeflow has been identified by proposed method incorrectly $\forall$ $h$, $s$, $d$ (FN)
\item $ Accuracy=\dfrac{TP+TN}{TP+FP+TP+TN}$
\end{itemize}

\begin{figure}[H]
 \centering
    \includegraphics[scale=0.55]{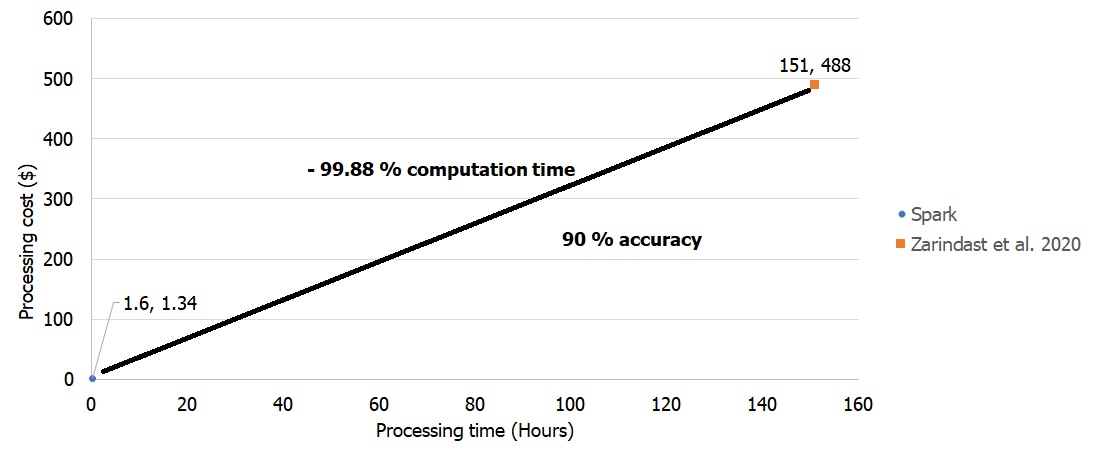}
    \caption{Cost, processing time, accuracy trade-off (Zarindast et al. (2020) vs proposed method). }%
    \label{fig:cost_accuracy trade off}%
\end{figure}

Based on a given computation power, we compared the estimated cost and processing time between the proposed framework and data-driven framework by \cite{a2}.
The processing cost estimation has been calculated using amazon web services (AWS) \cite{AWS} by giving computational power and time. These cost calculation are for processing INRIX 2018 probe data in our study location (see Figure \ref{fig:area.}) in interstates of Iowa, USA. 
The data-driven method proposed by zarindast et al. (2020) consists of three phases, namely 1- preprocessing 2-Bayesian change point detection, 3- RC, and NRC identification. Considering computation power given in table \ref{table:u}, the data-driven method took 151 hours of operation, which resulted in a total of  488 \$. While proposed framework Considering computation power given in table \ref{table:u} took 10 minutes, which resulted in 1.34 \$.
The accuracy of detection is based on detected congested hours for each date and segment between two algorithms. 
Figure \ref{fig:cost_accuracy trade off} shows the trade-off between computation time and accuracy. By only a 10 percent decrease in detection accuracy, we could decrease the computation time by 99.88 percent.
As shown in Figure \ref{fig:cost_accuracy trade off} proposed method was able to detect congestion with 84 percent accuracy as compared to the data-driven method proposed by zarindast et al. in (2020) \cite{a2}.

\section{CONCLUSIONS}
In this study, a highly parallelized framework for identifying freeways' congestion has been presented under RRD modeling. The proposed framework leverages a wealth of data to uncover Spatio-temporal congestion patterns further. We use state-of-the-art Spark as an efficient data processing system and HDFS cluster for computations; thus, the proposed framework is highly scalable and efficient in computation time. Most prior research is bound to a particular location or specific period. This study is among the first studies that consider a big data framework for identifying congestion and associated patterns to the best of our knowledge. The total input size was 76 GB, and the total computation time was 10 minutes. The proposed big data solution is a path toward real time traffic controlling and management. Comparison between the proposed method and data-driven definition proposed by Zarindast et al. in 2020 showed that the proposed method decreased computation time by 99.88 \% while maintained 90 \% congestion accuracy.
The experiment was conducted on Iowa's interstates, and the result is presented in terms of RC, delay cost and hours, congestion hours.

\printbibliography                                 

@preamble{ " \newcommand{\noop}[1]{} " }

@article{dean2008mapreduce,
  title={MapReduce: simplified data processing on large clusters},
  author={Dean, Jeffrey and Ghemawat, Sanjay},
  journal={Communications of the ACM},
  volume={51},
  number={1},
  pages={107--113},
  year={2008},
  publisher={ACM New York, NY, USA}
}

@online{storm,
  title = "Apache storm",
  year = {2015},
  url  ="https://storm.apache.org"
}

@article{abadi2016tensorflow,
  title={Tensorflow: Large-scale machine learning on heterogeneous distributed systems},
  author={Abadi, Mart{\'\i}n and Agarwal, Ashish and Barham, Paul and Brevdo, Eugene and Chen, Zhifeng and Citro, Craig and Corrado, Greg S and Davis, Andy and Dean, Jeffrey and Devin, Matthieu and others},
  
  journal={arXiv preprint arXiv:1603.04467},
  year={2016}
}

@article{tang2020survey,
  title={A Survey on Spark Ecosystem: Big Data Processing Infrastructure, Machine Learning, and Applications},
  author={Tang, Shanjiang and He, Bingsheng and Yu, Ce and Li, Yusen and Li, Kun},
  journal={IEEE Transactions on Knowledge and Data Engineering},
  year={2020},
  publisher={IEEE}
}

@article{zaharia2010spark,
  title={Spark: Cluster computing with working sets.},
  author={Zaharia, Matei and Chowdhury, Mosharaf and Franklin, Michael J and Shenker, Scott and Stoica, Ion and others},
  journal={HotCloud},
  volume={10},
  number={10-10},
  pages={95},
  year={2010}
}

@inproceedings{zaharia2012resilient,
  title={Resilient distributed datasets: A fault-tolerant abstraction for in-memory cluster computing},
  author={Zaharia, Matei and Chowdhury, Mosharaf and Das, Tathagata and Dave, Ankur and Ma, Justin and McCauly, Murphy and Franklin, Michael J and Shenker, Scott and Stoica, Ion},
  booktitle={Presented as part of the 9th $\{$USENIX$\}$ Symposium on Networked Systems Design and Implementation ($\{$NSDI$\}$ 12)},
  pages={15--28},
  year={2012}
}

@online{pyspark,
  title = "pyspark",
  year = {2016},
  url  ="http://spark.apache.org/docs/latest/api/python/index.htm"
}

@article{aljame2019dna,
  title={DNA short read alignment on apache spark},
  author={AlJame, Maryam and Ahmad, Imtiaz},
  journal={Applied Computing and Informatics},
  year={2019},
  publisher={Elsevier}
}

@inproceedings{seif2018stock,
  title={Stock market real time recommender model using apache spark framework},
  author={Seif, Mostafa Mohamed and Hamed, Essam M Ramzy and Hegazy, Abd El Fatah Abdel Ghfar},
  booktitle={International Conference on Advanced Machine Learning Technologies and Applications},
  pages={671--683},
  year={2018},
  organization={Springer}
}

@online{INRIX,
  title = "INRIX",
  year = {2018},
  url  ="http://inrix.com"
}

@inproceedings{an2011survey,
  title={A survey of intelligent transportation systems},
  author={An, Sheng-hai and Lee, Byung-Hyug and Shin, Dong-Ryeol},
  booktitle={2011 Third International Conference on Computational Intelligence, Communication Systems and Networks},
  pages={332--337},
  year={2011},
  organization={IEEE}
}

@article{rehborn2011empirical,
  title={An empirical study of common traffic congestion features based on traffic data measured in the USA, the UK, and Germany},
  author={Rehborn, Hubert and Klenov, Sergey L and Palmer, Jochen},
  journal={Physica A: Statistical Mechanics and its Applications},
  volume={390},
  number={23-24},
  pages={4466--4485},
  year={2011},
  publisher={Elsevier}
}

@article{levy2010evaluation,
  title={Evaluation of the public health impacts of traffic congestion: a health risk assessment},
  author={Levy, Jonathan I and Buonocore, Jonathan J and Von Stackelberg, Katherine},
  journal={Environmental health},
  volume={9},
  number={1},
  pages={65},
  year={2010},
  publisher={Springer}
}

@article{zhao2019geographical,
  title={Geographical patterns of traffic congestion in growing megacities: Big data analytics from Beijing},
  author={Zhao, Pengjun and Hu, Haoyu},
  journal={Cities},
  volume={92},
  pages={164--174},
  year={2019},
  publisher={Elsevier}
}

@misc{atousa:online,
author = {Atousa, Zarindast},
title = {A data driven method for congestion mining using big data analytic},
howpublished = {\url{https://lib.dr.iastate.edu/creativecomponents/441/}},
month = {},
year = {2019}

}

@article{a2,
  title={A data-driven method for congestion identification and classification},
  author={Atousa, Zarindast and Poddar,Suphadipto  and Anuj , Sharma},
  journal={Expert system with application},
  volume={},
  pages={},
  year={\noop{2020}submitted},
  publisher={Elsevier}
}

@online{AWS,
  title = "AWS",
  year = {2018},
url= "https://calculator.s3.amazonaws.com/index.html#r=IAD&key=files/calc-57352e003337796e7e8ec1a8ce1a5f274428dd7b&v=ver20200317o8"
}

@article{zarindast2018determination,
  title={Determination of Bus Station Locations under Emission and Social Cost Constraints},
  author={Zarindast, Atousa and G{\"u}nay, Elif El{\c{c}}in and Park, Kijung and Okudan Kremer, G{\"u}l E},
  year={2018}
}

@article{gunay2020multi,
  title={A multi-objective robust possibilistic programming approach to sustainable public transportation network design},
  author={G{\"u}nay, Elif El{\c{c}}in and Kremer, G{\"u}l E Okudan and Zarindast, Atousa},
  journal={Fuzzy Sets and Systems},
  year={2020},
  publisher={Elsevier}
}

@article{zarindast2019data,
  title={A data driven method for congestion mining using big data analytic},
  author={Zarindast, Atousa},
  year={2019}
}

@article{shahhosseini2020improved,
  title={Improved Weighted Random Forest for Classification Problems},
  author={Shahhosseini, Mohsen and Hu, Guiping},
  journal={arXiv preprint arXiv:2009.00534},
  year={2020}
}

@article{rajabalizadeh2020exploratory,
  title={An Exploratory Analysis of Electronic Intensive Care Unit (eICU) Collaborative Research Database},
  author={Rajabalizadeh, Atefeh and Norouzi Nia, Javad and Safaei, Nima and Talafidaryani, Mojtaba and Bijari, Reyhaneh and Zarindast, Atousa and Fotouhi, Fateme and Salehi, Masud and Moqri, Mahdi},
  year={2020}
}
\end{document}